\documentclass[aps,prl,twocolumn,showpacs,floatfix]{revtex4}

\usepackage[pdfview=FitV,pdfstartview=FitV]{hyperref}   % use for hypertext links, including those to external documents and URLs
\usepackage{amsmath}
\usepackage{amssymb}    % need for subequations
\usepackage{graphicx}   % need for figures
\usepackage{epsfig}
\usepackage{color}

\newcommand{\hn}{{\hat n}}
\newcommand{\J}{ {{\cal J}_0} }

\begin{document}

%opening
\title{Ultracold Lattice Gases with Periodically Modulated Interactions}
\author{\'Akos Rapp, Xiaolong Deng, and Luis Santos}
\affiliation{ Institut f\"ur Theoretische Physik, Leibniz Universit\"at, 30167 Hannover, Germany}

\date{\today}

\begin{abstract}
We show that a time-dependent magnetic field inducing a periodically modulated scattering length may 
lead to interesting novel scenarios for cold gases in optical lattices, characterized by a
nonlinear hopping depending on the number difference at neighboring sites. 
We discuss the rich physics introduced by this hopping, including pair superfluidity, 
exactly defect-free Mott-insulator states for finite hopping, and pure holon and doublon superfluids. 
We also address experimental detection, showing that the introduced nonlinear hopping may lead 
in harmonically trapped gases to abrupt drops in the density profile marking the interface between 
different superfluid regions.
\end{abstract}

% \DOI{10.1103/PhysRevLett.109.203005}


\maketitle

% Introduction

Ultracold atoms in optical lattices formed by laser beams
provide an excellent environment for studying lattice models of general
relevance in condensed-matter physics, and in particular, variations of the
celebrated Hubbard model~\cite{BlochZwergerDalibard,Review-Lewenstein}.
Cold lattice gases allow for an unprecedented degree of control of various experimental
parameters, even in real time. In particular, interparticle interactions can be changed
by means of Feshbach resonances~\cite{Review-Feshbach}.
Moreover, recent milestone achievements allow for 
site-resolved detection, permitting the study of \textit{in situ} densities~\cite{Bakr2010,Sherson2010}, 
and more involved measurements, as that of nonlocal parity order~\cite{Bloch-parity}.

The modulation of the lattice parameters in real time opens interesting possibilities of control and
quantum engineering. In particular, a periodic lattice shaking translates by means of Floquet's theorem~\cite{Floquet0,Floquet-Haengii} into a modified hopping constant~\cite{Floquet2}, which may even reverse its sign as shown in experiments~\cite{latticeshaking,Kierig2008}. This technique has been employed to drive the Mott-insulator (MI) to superfluid (SF) transition~\cite{Zenesini2009}, and to simulate frustrated classical magnetism~\cite{Sengstock}.
Recent experiments have explored as well the fascinating perspectives offered by periodically driven lattices in strongly correlated gases~\cite{Chen2011,Ma2011}.

% Non-linear hopping: Dipoles, Multiband physics
The effective Hubbard-like models describing ultracold lattice gases are typically characterized by 
a hopping independent of the number of particles at the sites. This is, however, not necessarily 
the case. Multiband physics~\cite{PSF4,Frank,occdep_params} and dipolar interactions for sufficiently large dipole moments~\cite{PSF5} 
may lead to occupation-dependent hopping. A major consequence of nonlinear hopping  is the possibility to observe pair superfluidity~(PSF)~\cite{PSF1,PSF5}, which resembles pairing in SF Fermi gases, although for bosons superfluidity exists as well without pairing.

% This Letter
In this Letter, we consider a cold lattice gas in the presence of a periodically modulated magnetic field.  
In the vicinity of a Feshbach resonance, this field induces modulated interparticle 
interactions~\cite{as-modulation-exp}. 
Interestingly, Ref.~\cite{Gong2009} has shown that periodic modulations of the interaction strength may lead to a many-body coherent destruction of tunneling in two-mode Bose-Einstein condensates.
As shown below, the generalization of this effect to lattice gases leads under proper conditions 
to an effective Hubbard-like model with a non-linear hopping which, in contrast to other proposals mentioned above,  
depends on the difference of occupations at neighboring sites, and retains its nonlinear character even for 
weak lattices. We discuss the rich physics introduced by this hopping, including pair superfluid phases, 
exactly defect-free MI states for finite hopping, and pure holon and doublon superfluids. We also address
experimental detection, showing that the studied nonlinear hopping may lead to abrupt drops in the density profile of harmonically trapped gases.

% Model

We consider bosons in a lattice in the presence of a periodically modulated magnetic field $B(t)=B(t+T)$ (with period $T=2\pi/\omega$) 
chosen close to a Feshbach resonance, where the $s$-wave scattering length acquires the form 
$a(t)=a_{bg}\left ( 1 - \frac{\Delta B}{B(t)-B_r} \right )=a_0+\sum_{l>0} a_l \cos(l\omega t)$.
Here $\Delta B$ and $B_r$ determine the width and position of the resonance, respectively, and $a_{bg}$ is the background scattering length~\cite{Review-Feshbach}.
Assuming that the gap between the first two lattice bands is much larger than any other energy 
scale in the problem, we consider only the lowest band and describe the system
by a Bose-Hubbard model~(BHM)~\cite{BlochZwergerDalibard,Review-Lewenstein}:
\begin{eqnarray}
 H(t) = - J\sum_{\langle ij\rangle} b_i^\dag b_j  + \frac{U(t)}{2}\sum_i 
\hn_i \left ( \hn_i -1\right )- \sum_i \mu \; \hn_i,
  \label{eq:def:H(t)}
\end{eqnarray}
where $b_i$ ($b_i^\dag$) is the bosonic annihilation (creation) operator at site $i$, $\hn_i =b_i^\dag b_i$, 
$\mu$ is the chemical potential, $J>0$ is the hopping rate and 
$\langle..\rangle$ denotes nearest neighbors. Interactions are characterized by a
coupling $U(t)=U_0+\sum_{l>0} U_l \cos (l\omega t)=U_0+{\tilde U}(t)$, with $U_0>0$ and $U_l=\frac{4\pi\hbar^2 a_l}{M} \int d^3 r |w({\bf r})|^4$. Here
$w({\bf r})$ is the lowest Wannier function and $M$ is the atomic mass.

% Floquet analysis: effective assisted hopping
We apply a similar analysis as the one used for shaken lattices~\cite{Floquet2}. 
We specify a Floquet basis
\begin{equation}
 |\{n_j\},m\rangle =e^{im\omega t} e^{- i\frac{V(t)}{2}\sum_j\hn_j \left ( \hn_j -1\right )} |\{n_j\} \rangle \;,
\label{eq:Floquet}
\end{equation}
where $m$ defines the Floquet sectors and $|\{n_j\} \rangle$ is the Fock basis, characterized by the atom 
number at each site. In Eq.~\eqref{eq:Floquet}, we defined
$V(t)=\int^t\!{\tilde U}(t')dt'\!/\!\hbar$. We introduce the time-averaged scalar product 
$\langle\langle \{n'_j\},m' | \dots |\{n_j\},m \rangle\rangle =\frac{1}{T}\!\int_0^{T}\!\langle \{n'_j\},m'| \dots |\{n_j\},m\rangle$, and establish 
the matrix elements:
\begin{eqnarray}
&&\!\!\!\!\!\!\!\!\!\langle\langle \{n'_j\},m' | [H(t)-i\hbar\partial_t]  |\{n_j\},m\rangle\rangle  \nonumber \\
&& = \delta_{m,m'} \; \langle \{ n'_j\}|  H_{m}  \vert \{ n_j\} \rangle \nonumber \\
&& \phantom{=}-J\sum_{\langle i,j\rangle}  \langle \{ n'_j\}| b_i^\dag F_{m'-m}(\hn_i-\hn_j) b_j  \vert \{ n_j\} \rangle,  
\end{eqnarray}
with $H_m=m\hbar\omega+\frac{U_0}{2}\sum_j\hn_j \left ( \hn_j -1\right )-\sum_j \mu \hn_j$
and $F_{m}(x)=\frac{1}{T}\int_0^T dt e^{-imt}e^{iV(t)x}$.
If $\hbar\omega\gg J,U_0$, we may restrict to a single Floquet sector $m=0$, resulting in an effective
time-independent Hamiltonian of the form
\begin{eqnarray}
 H_{\rm eff} &=& - J\sum_{\langle ij\rangle} b_i^\dag F_{0}(\hn_i-\hn_j) b_j  \nonumber \\
 &+& \frac{U_0}{2}\sum_i  \hn_i \left ( \hn_i -1\right )- \mu\sum_i \hn_i .
  \label{eq:def:Heff(t)}
\end{eqnarray}
Hence, interactions with a periodic modulation result in a 
nonlinear hopping term, which depends on the atom number difference between neighboring sites.
Note that this  nonlinear character remains relevant for any value of the bare hopping $J$.
In the following we discuss the specific case ${\tilde U}(t)=U_1 \cos\omega t$. In this case,
$F_0(x)=\J(\Omega x)$, with $\J$ the Bessel function and $\Omega=U_1/\hbar\omega$, generalizing the result
of Ref.~\cite{Gong2009} for two-well Bose-Einstein condensates. 

%%%%%
%% FIGURE 1
\begin{figure}[t]
 \centering
\includegraphics[clip=true,width=0.45\textwidth]{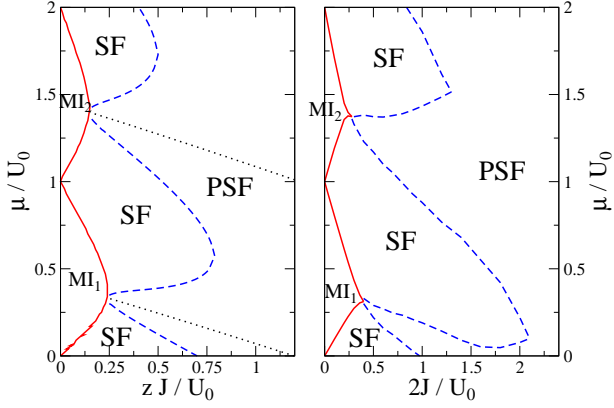}
\vspace*{-0.2cm}
 \caption{ (color online). Phase diagram for $\Omega=4$ using the GA~(left) and the DMRG~(right). 
Solid curves define the MI lobes, whereas dashed curves are the SF-PSF boundaries. 
The black dotted line indicates $\langle b_i \rangle = 0$.
}
\label{fig:1}
\end{figure}
%%%%%

% Gutzwiller Ansatz

Insight into Eq.~\eqref{eq:def:Heff(t)} is gained by means of a 
Gutzwiller ansatz~(GA) for the ground state~\cite{B-Gutzwiller}, 
$ | G \rangle = \prod_j \sum_{n} f_n(j) | n_j \rangle$, 
where $f_n(j)$ are variational parameters [$\sum_n |f_n(j)|^2=1$], 
determined by minimizing $\langle G | H_{\rm eff} | G \rangle$.
Results obtained by choosing homogeneous real $f_n(j)= f_n$~\cite{footnote0}  
are shown in Fig.~\ref{fig:1}~(left), where we depict the 
mean-field phase diagram for $\Omega=4$ [$\J (\Omega)\simeq -0.4$]
as a function of $\mu/U_0$ and $zJ/U_0$, with $z$ the coordination number. 
As usual, MI phases are characterized by integer $\langle \hat n_i \rangle $, and vanishing single-particle- and pair-condensation fractions, $\rho_1\equiv |\langle b_i\rangle|^2/\langle \hat n_i \rangle$ and 
$\rho_2=|\langle b_i^2\rangle|^2/\langle \hat n_i \rangle^2$. 
The superfluid regime may be split into two different phases separated by a crossover: a usual superfluid,  
characterized by a dominant single-particle condensation, $\rho_1>\rho_2>0$, 
and a pair superfluid, where pair-condensation dominates, $\rho_2>\rho_1\geq 0$.
Pair superfluidity is especially pronounced in the vicinity of integer $\langle \hat n_i \rangle$, 
where $\langle b_i \rangle = 0$. Our GA results show that PSF only occurs if $\J(\Omega)<0$. 
This may be understood by considering integer $\langle \hat n \rangle=n$, 
and restricting the variational space to $f_{n\pm 1}=\frac{\sin\eta}{\sqrt{2}}e^{i\varphi_{\pm }}$ and $f_{n}=\cos\eta$. 
For $\J(\Omega)<0$, energy minimization gives  $\bar\varphi\equiv\varphi_+ + \varphi_-=\pi$, while 
$\bar\varphi=0$ for $\J(\Omega)>0$. As a result, for $2Jz/U_0\gg 1$, 
PSF demands $(n+1)>2(\sqrt{n}+{\rm sgn} (\J(\Omega))\sqrt{n+1})^2$, which is only fulfilled if $\J(\Omega)<0$.

% 1D

% Since GA is not reliable in 1D, 
To complement the mean-field GA results, we have also employed numerically exact methods in one dimension. In particular, we used the density-matrix renormalization group~(DMRG)~\cite{DMRG} with up to $40$ sites and keeping $200$ states, and a related method, the infinite time-evolving block 
decimation~(iTEBD) method~\cite{iTEBD} using a Schmidt dimension of $200$. We have monitored the behavior of single-particle and pair correlations, $G_1(i,j)\equiv \langle b_i^\dag b_j\rangle $ and $G_2(i,j)\equiv \langle ( b_i^\dag )^2 b_j^2 \rangle $, respectively. Both decay exponentially in the Mott insulator. 
For both the PSF and SF regions, both $G_{1}$ and $G_2$ have a power-law decay~\cite{footnote4}, but in the PSF $G_2$ decays slower than $G_1$. The opposite characterizes the SF phase. Figure~\ref{fig:1}~(right) shows the one-dimensional phase diagram for $\Omega=4$, which closely resembles the one obtained using GA. Similar to the GA, we observe a pair superfluid phase, which for integer $\langle \hat n\rangle$ approaches all the way to the tip of the MI lobes. Away from the lobe tips we observe a direct MI-SF transition. 
Our one-dimensional results also confirm the absence of PSF for $\J(\Omega)>0$.

% Case of J0(Omega)=0

% No particle-hole creation/destruction
The case $\J(\Omega)=0$ is particularly interesting, since for neighboring sites $i$ and $j$ with equal 
number of particles, the process $|n\rangle_i |n\rangle_j \rightarrow |n\pm 1\rangle_i |n\mp 1\rangle_j$ is forbidden.
However, the hopping $|n\pm 1\rangle_i |n\rangle_j \rightarrow |n\rangle_i |n\pm 1\rangle_j$ is still characterized by the usual 
rate $J$. This difference has a remarkable impact for both the MI and the SF phases.

\begin{figure}[t]
\centering
\includegraphics[clip=true,width=0.38\textwidth]{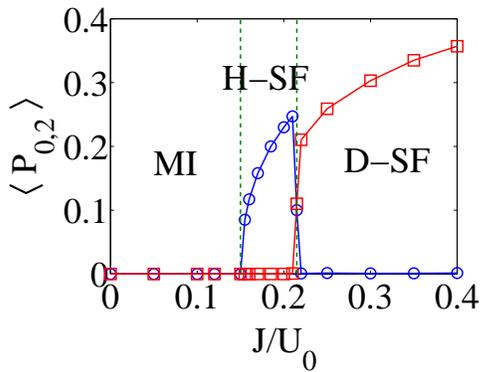}
\vspace*{-0.2cm}
\caption{(color online). iTEBD results for the holon~(circles) and doublon~(squares) populations 
as a function of $J/U_0$ for a one-dimensional system with $\mu/U_0=0.3$ and $\Omega=2.405$. Note the absence of defects in the Mott insulator ($J/U_0< 0.15$), and the appearance of the holon SF (H-SF) and doublon SF (D-SF).
}
\label{fig:2}
\end{figure}

% Mott insulator phases with J(Omega)=0
For $J=0$, the ground state of Eq.~\eqref{eq:def:Heff(t)} is, as for the standard BHM~($\Omega=0$), 
a defect-free MI $\bigotimes_j |n\rangle_j$ for $n-1<\mu/U_0<n$~\cite{FisherFisher}.   
For $\Omega=0$ and $J>0$, this state is not an eigenstate of Eq.~\eqref{eq:def:Heff(t)}, 
and quantum fluctuations induce a finite particle-hole population in the Mott insulator with an associated 
nonlocal parity order~\cite{Bloch-parity}. Interestingly, the defect-free state remains an eigenstate 
of Eq.~\eqref{eq:def:Heff(t)} for $\J(\Omega)=0$. As a result, the whole MI lobe is characterized 
by the absence of particle-hole defects. 
Although this is typically an artifact in the mean-field GA,
in this case it is an exact result for any dimensions. This is illustrated for the one-dimensional case 
in Fig.~\ref{fig:2}, where our iTEBD results show a vanishing variance $(\Delta n)^2=\langle \hat n^2\rangle - \langle \hat n\rangle^2$ within the whole Mott insulator. 
A defect-free Mott insulator may be revealed by parity-order measurements 
using site-resolved techniques~\cite{Bloch-parity}. Whereas for the standard BHM doublon-hole pairs 
reduce parity order, a defect-free Mott insulator results in unit parity in the whole Mott phase.

%%%%%
%% FIGURE 2

% Superfluid phases
Conversely, particles or holes~($|n\pm 1\rangle$) on top of the state $\bigotimes_j |n\rangle_j$ acquire also remarkable properties. For any $\Omega$, extra particles and holes move with a hopping rate $(n+1)J$ and $nJ$, respectively. 
For $\Omega=0$ defects are unstable, being created and destroyed by processes 
$|n\rangle_i |n\rangle_j \leftrightarrow |n\pm 1\rangle_i |n\mp 1\rangle_j$. Since these processes are forbidden 
for $\J(\Omega)=0$, defects remain stable~\cite{footnote1}. Neglecting 
occupations other than $n$ and $n\pm 1$, the defects are 
described by an effective Hamiltonian $H_h+H_p$, where 
\begin{eqnarray}
H_{h}&=&-Jn\!\!\sum_{<i,j>}\! h_i^\dag h_j +(\mu-U_0(n-1))\sum_i  h_i^\dag h_i, \label{eq:Hh}\\
H_{p}&=&-J(n+1)\!\!\sum_{<i,j>}\! p_i^\dag p_j +(U_0n-\mu)\sum_i  p_i^\dag p_i, \label{eq:Hp} 
\end{eqnarray}
characterize, respectively, the physics of holes and particles, with the hard-core assumption 
$p_i^\dag p_i + h_i^\dag h_i =0$ or $1$, with $h_i$~($p_i$) the %bosonic 
operators for extra 
holes~(particles) at site $i$. In Eqs.~\eqref{eq:Hh} and~\eqref{eq:Hp}  we have set the energy of the defect-free MI state 
$E_{\rm MI}=0$. Thus the system behaves as a two-component hard-core lattice Bose gas.
For higher dimensions, a dilute gas of extra holes (holon gas) may be considered as a basically 
free (superfluid) Bose gas, with a dispersion $E_h({\bf q})=\mu-U_0(n-1)+ n \epsilon^0_{\bf q}$, where $\epsilon^0_{\bf q} = -2J \sum_{j=x,y,z}\cos(q_j d)$ for a three-dimensional cubic lattice and $d$ is the lattice spacing. On the other hand, the dilute gas
of extra particles (``doublon'' gas~\cite{footnote2}) has a dispersion 
$E_p({\bf q})=U_0n-\mu + (n+1)\epsilon^0_{\bf q}$. 

\begin{figure}[t]
\centering
\vspace*{-0.2cm}
\includegraphics[clip=true,width=0.40\textwidth, trim=0 0 0 80]{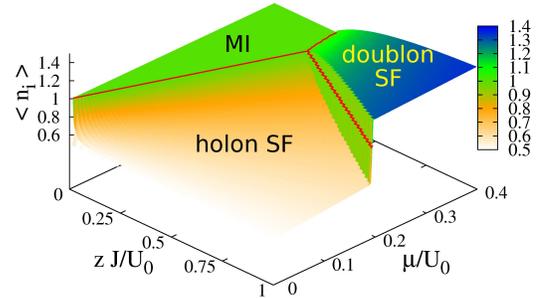}
\vspace*{-0.3cm}
\caption{(color online). Homogeneous GA results for $\langle \hat n \rangle$ as a function of $J/U_0$ and $\mu/U_0$ for $\Omega=2.405$. 
The solid red lines denote the boundary of the Mott insulator and the line of integer filling $1$. Note the abrupt jump in the density at that line, indicating the 
transition between the holon SF and doublon SF regimes.}
\label{fig:3}
\end{figure}

At zero temperature, the defect gas condenses for $\mu<\mu_c\equiv U_0(n-1/2)-Jz$
at the bottom of the holon band, $E_h(0)$, acquiring a pure holon character. On the other hand, for $\mu>\mu_c$ the system condenses at $E_p(0)$ into a pure doublon gas. Hence, remarkably, we expect an abrupt jump 
of $\langle \hat n \rangle$~(i.e. a diverging compressibility) at the line $\mu=\mu_c$, which coincides with the line of integer $\langle \hat n \rangle =n$. 
Figure~\ref{fig:3} depicts our GA results for the density as a function of $\mu/U_0$ and $J/U_0$, 
which, as expected from the previous discussion, presents an abrupt jump between a holon and a doublon superfluid. 

%%%%%
%% FIGURE 3

In one dimension, the defects behave, due to the hard-core constraint, rather as a two-component Tonks 
gas, but a similar two-band reasoning applies, and we may also expect the existence of 
pure holon and doublon superfluids. Figure~\ref{fig:2} shows our iTEBD results in the vicinity of $\langle \hat n  \rangle =1$ for the holon~(doublon) 
populations $\langle \hat P_0 \rangle $~($\langle \hat P_2 \rangle$) , with $\hat P_{n}=\prod_{n'\neq n} (\hat n-n')/ (n-n')$. In addition to the Mott insulator  
characterized by $\langle\hat  P_{0,2} \rangle=0$, we observe a holon SF
($\langle \hat P_2 \rangle=0$) and an abrupt jump to a doublon SF~($\langle\hat  P_0 \rangle=0$).
Note that a pure doublon SF or holon SF excludes PSF. 

At constant $\mu$ for $\J(\Omega) = 0$ the system undergoes a MI - doublon (holon) SF transition at a critical tunneling $J_c(\mu)$ for which $E_{p(h)}(0)=E_{\rm MI}$. 
On the contrary, at constant integer $\langle \hat n \rangle$, 
there is no one-dimensional MI-SF transition at finite hopping $J$. 
Due to the absence of processes $|n\rangle_i |n\rangle_j \leftrightarrow |n\pm 1\rangle_i |n\mp 1\rangle_j$, 
doublons and holons cannot swap their positions through second-order superexchange. As a result, if holons and doublons coexist (which only happens at the singular integer filling line), superfluidity is absent. Our DMRG results for $\langle \hat n \rangle=1$ confirm this insulating character for any $J$, showing a clear transition between a defect-free insulator and an insulator with a finite defect density.
 
For a finite but small $\J(\Omega)$, the SF regions retain to a large extent their holon and doublon character, although the concentration of doublons in the holon SF and holons in the doublon SF increases for growing $\J(\Omega)$ and $J$. 
The coexistence region for holons and doublons is hence not any more singular, 
although it remains characterized by a large compressibility for small $\J (\Omega)$. 
For $\J(\Omega)<0$ this coexistence region becomes the pair superfluid phase discussed above. Away from the Mott tip 
a direct MI-SF transition is observed, as discussed above, since at the MI boundary holons and doublons do not coexist.

% Experiments

Let us finally discuss some experimental questions. 
Optimal experimental conditions for periodically modulated interactions are provided by $^{85}$Rb, which has a particularly large
$a_{bg}\simeq -400a_B$ (with $a_B$ the Bohr radius), and a broad Feshbach resonance at $B_{r}=155.2$G, with a width $\Delta B=11.6$G~\cite{Bzero-85Rb}. 
The desired form $a(t) \approx a_0+a_1 \cos(\omega t)$ can be achieved for a magnetic field dependence $B(t)/G\simeq 167.56+5.58 \cos(\omega t)$, with  $a_0\approx20a_B$ and $a_1\approx 200a_B$. We consider a lattice spacing $d=0.5\mu$m, and potential depth $V_L=sE_R$, where $E_R=\hbar^2\pi^2/2Md^2$ is the recoil energy. For $s \approx 17$ ($J \ll U_0$), the value $\Omega=2.4$~($\J(\Omega)\simeq 0$) is obtained for $\omega\simeq 2\pi\times 900$ Hz $\gg U_0/\hbar=2\pi\times 217$ Hz, ensuring that only one Floquet manifold is relevant~\cite{footnote3}.

\begin{figure}[t]
\centering
\includegraphics[clip=true,width=0.38\textwidth]{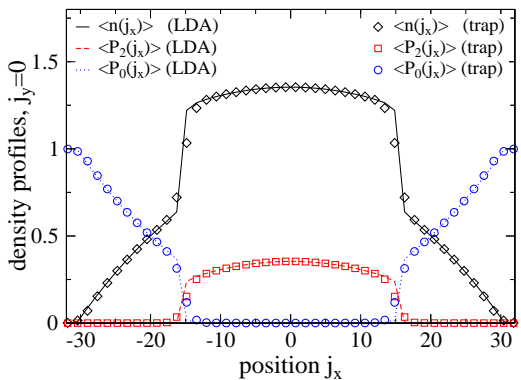}
\vspace*{-0.2cm}
\caption{\label{fig:fixcuts} (color online). GA results for the site densities $\langle \hat n_j \rangle$ and the on-site holon and doublon populations for a two-dimensional lattice with a harmonic confinement $V(j_x,j_y)=V_0 (j_x^2+j_y^2)$, with $V_0/J=0.0075$, interaction $U_0/J=5.33$, a central chemical potential $\mu_0/J=3$, and $\Omega=2.45$~($\J(\Omega)\simeq 0$). Note the central doublon-SF region, surrounded by a holon-SF ring, and the abrupt density drop separating both regimes. Lines indicate local-density approximation~(LDA) results.}
\label{fig:4}
\end{figure}

In order to address the question of detection, we have to consider the transformation [Eq.~\eqref{eq:Floquet}] between the Floquet $|\{n_j\},m\rangle$ and the Fock $|\{n_j\}\rangle$ basis. The densities $\langle \hat n_i\rangle$ are 
equivalent in both; therefore, the large compressibility regions characteristic of 
$|\J(\Omega)|\simeq 0$ may be revealed in \textit{in situ} experiments with an additional harmonic confinement. This is illustrated in Fig.~\ref{fig:4}, where we show inhomogeneous GA results for a harmonic trap in two dimensions. As expected from the local-density approximation, 
we observe an abrupt density jump when the local chemical potential crosses its critical value. 

%%%%%
%% FIGURE 4

Interpretation of other observables, as e.g. the momentum distribution in time-of-flight~(TOF) measurements, 
may be more involved, since 
$\langle \{n'_j\}| b_i^\dag b_j |\{n_j\}\rangle \sim e^{-iV(t)(n_i-n_j+1)}\langle \{n'_j\},m| b_i^\dag b_j |\{n_j\},m\rangle$. 
However, for the holon SF and doublon SF phases the TOF measurement is almost 
time-independent for small $|\J(\Omega)|$. Indeed this weak dependence is in itself a proof of the holon or doublon character of the SF. For 
large $|\J(\Omega)|$ the nonlinear conversion is an issue, and in general measurement results are periodic.

In summary, periodically modulated interactions lead to a rich physics for cold gases in optical lattices, characterized by a nonlinear hopping depending on the number difference at neighboring sites. This hopping can lead to pair superfluid phases, and also to defect-free Mott states, and holon and doublon superfluids, which may be revealed by parity measurements and by abrupt jumps of the \textit{in situ} densities in harmonically trapped lattice gases.

% \acknowledgements 
We acknowledge financial support by the Cluster of Excellence QUEST.


\begin{thebibliography}{99}
\bibitem{footnote0} Allowing different $f_n$ on the two sublattices 
	of a bipartite lattice did not lead to variational improvement.
\bibitem{footnote4} For integer $\langle \hat n \rangle$ $G_1$ decays exponentially, i.e.,
	the phase is purely pair superfluid(in agreement with GA results).
\bibitem{footnote1} Processes such as
	$|n+2\rangle_i |n-1\rangle_j \leftrightarrow |n+1\rangle_i |n\rangle_j$ may still occur, but for 
	$\langle \hat n \rangle$ in the vicinity of $n$ they are much less probable, especially at low hopping and low $n$.
\bibitem{footnote2} The term ``doublon'' is used here as a general concept for extra particles on top of a MI at any integer $\langle\hat n\rangle$.
\bibitem{footnote3} Parametric resonances to upper Bloch bands should be avoided. For the discussed example
these resonances are avoided since the gap to the second band is $\simeq 17.5 \hbar\omega$.
\end{thebibliography}
\end{document}